\documentclass{jpsj2}

\newcommand{\Tc}{$T_\mathrm{c}$}
\newcommand{\LaFePO}{LaFePO$_{1-x}$F$_{x}$}
\newcommand{\LaFeAsO}{LaFeAsO$_{1-x}$F$_{x}$}
\newcommand{\R}{\textit{R}}

\title{Systematic Study on Fluorine-doping Dependence of Superconducting and Normal State Properties in LaFePO$_{1-x}$F$_{x}$}

\author{Shinnosuke \textsc{Suzuki}$^{1}$, Shigeki \textsc{Miyasaka}$^{1,2}$\thanks{E-mail: miyasaka@phys.sci.osaka-u.ac.jp}, Setsuko \textsc{Tajima}$^{1,2}$, Takanori \textsc{Kida}$^{2,3}$, and Masayuki \textsc{Hagiwara}$^{2,3}$}

\inst{$^{1}$Department of Physics, Graduate School of Science, Osaka University, Toyonaka, Osaka 560-0043 \\
$^{2}$JST, TRIP, Chiyoda, Tokyo 102-0075 \\
$^{3}$Center for Quantum Science and Technology under Extreme Conditions, Osaka University, Toyonaka, Osaka 560-8531}

\abst{We have investigated the fluorine-doping dependence of lattice constants, transports and specific heat for polycrystalline LaFePO$_{1-x}$F$_{x}$. F doping slightly and monotonically decreases the in-plane lattice parameter. In the normal state, electrical resistivity at low temperature is proportional to the square of temperature and the electronic specific heat coefficient has large value, indicating the existence of moderate electron-electron correlation in this system. Hall coefficient has large magnitude, and shows large temperature dependence, indicating the low carrier density and multiple carriers in this system. Temperature dependence of the upper critical field suggests that the system is a two gap superconductor. The F-doping dependence of these properties in this system are very weak, while in the FeAs system (LaFeAsO), the F doping induces the large changes in electronic properties. This difference is probably due to the different F-doping dependence of the lattice in these two systems. It has been revealed that a pure effect of electron doping on electronic properties is very weak in this Fe pnictide compound.}

\kword{superconductivity, iron-based superconductor, LaFePO}

\begin{document}
\maketitle
%%%%%%%%%%%%%%%%%%%%%%%%%%%%%%%%%%%%%%%%%%%%%%%%%%%%%
\section{Introduction}

Recently, superconductivity in iron- and nickel-based oxyphosphides, LaFePO and LaNiPO were reported.~\cite{kamihara1,watanabe} In the same family of iron-based oxypnictides, Kamihara and co-workers have found a new superconductor, {\LaFeAsO} with critical transition temperature, {\Tc} = 26 K.~\cite{kamihara0} Since then, the research on the iron-based pnictides have been active, and the superconductivity with higher {\Tc} above 50 K has been discovered in F-doped and O-deficient FeAs systems with other rare-earth elements ($R$), {\R}FeAsO$_{1-x}$F$_{x}$ and $R$FeAsO$_{1-\delta}$.~\cite{chen1,ren1,chen2,kito} Much experimental and theoretical efforts are now being made to clarify the superconducting mechanism in these Fe-based pnictide systems. \par

These FeP and FeAs systems are composed of alternant stacked Fe-P/As and {\R}-O layers. 
The substitution of F$^{-}$ for O$^{2-}$ or the oxygen deficiency in the {\R}-O layers introduces the charge carriers (electrons) to the conductive layers of Fe-P/As. 
The isostructural FeP and FeAs systems have a similar electronic structure.~\cite{vildosola} In addition, the theoretical investigation has indicated that the bands around Fermi level have mainly Fe 3$d$ character, while the $p$ bands of P and As are quite minor in these bands. Nevertheless, the FeP and FeAs systems show distinct differences in the superconducting, electronic and magnetic properties.~\cite{kamihara1,kamihara0,cruz,klauss,huang,luetkens} 
This is still an open question. 
In order to understand the superconductivity in the FeAs compounds and the origin of difference in the superconducting/normal state properties between these FeP and FeAs systems, it is very important to clarify the electronic properties in {\LaFePO}. \par

LaFePO, which is the end member of the FeP system, shows a superconducting behavior below {\Tc} $\sim$ 4 K, and becomes a paramagnetic metal above {\Tc}.~\cite{kamihara1} It shows no structural phase transition with decreasing temperature, keeping a tetragonal lattice. So far, few research groups have reported the superconducting properties in the F-doped FeP compound, because of the difficulty of their synthesis. In refs. [1] and [13], Y. Kamihara $et$ $al.$ have reported only 6$\%$ F-doped sample. In the present work, we have synthesized several F-doped FeP compounds and systematically investigated the F-doping dependence of lattice parameters, transports and specific heat of {\LaFePO} to clarify the structural and electronic features in this system. \par

%%%%%%%%%%%%%%%%%%%%%%%%%%%%%%%%%%%%%%%%%%%%%%%%%%%%%%
\section{Experimental procedures}
Polycrystalline samples of {\LaFePO} ($x$ = 0-0.10) were synthesized by using a two-step solid-state-reaction method. First, LaP powder was synthesized by heating a mixture of La and P powders in an evacuated silica tube at 400 $^\circ$C for 12 hours and 700 $^\circ$C for 12 hours. Second, Fe, dehydrated Fe$_2$O$_3$, LaF$_3$ and P powders were mixed with the prepared LaP and pressed into pellets under a pressure of 10 MPa. Since the starting materials seem to be partially oxidized~\cite{liang,kondrat}, 3 at.\% of oxygen was reduced from the stoichiometry by changing the ratio of Fe and Fe$_2$O$_3$ powders. The pellets were annealed in evacuated silica tubes at 1100 $^\circ$C for 40 hours. The second step was carried out under a nitrogen atmosphere to prevent the LaP from reacting with oxygen and moisture.\par
The samples of LaFePO$_{1-x-y}$F$_x$ with several oxygen deficiency $y$ have been synthesized by using the above-mentioned method. In the samples with $y<$0.03, the several impurities such as La$_2$O$_3$, LaOF and so on, were observed by using powder X-ray diffraction. In contrast, the powder X-ray diffraction patterns for $y \ge$0.03 indicate the samples are a single phase. In the last step of synthesis, only the samples with the pellet shape were annealed in the evacuated silica tubes. So, we concluded that the samples with the nominal molecular formula of LaFePO$_{0.97-x}$F$_x$ have almost nonstoichiometry of oxygen and fluorine in the present work.\par
The samples were ground and checked by powder X-ray diffraction using Cu $K_{\alpha_1}$ radiation at room temperature.
All diffraction peaks of respective samples can be assigned to the calculated Bragg peaks for the tetragonal LaFePO~\cite{zimmer}. The in-plane ($a$) and out-of-plane lattice constants ($c$) of {\LaFePO} with F concentrations of $x$= 0-0.10 are obtained by using least squares fitting of the measured peak positions of the powder X-ray diffraction data in $2\theta$ = 20 $^\circ$ - 100 $^\circ$. \par
The magnetic susceptibility were measured in a magnetic field of 10 Oe to observe the diamagnetic behavior due to the superconductivity and determine {\Tc} for the {\LaFePO} samples. The volume fractions of the superconducting phase estimated from the diamagnetic susceptibility at 2 K are over 80 $\%$ for all {\LaFePO} samples. In this paper, the critical temperatures {\Tc} are defined by an onset transition temperatures observed in the temperature-dependent susceptibility. \par

Transport and specific heat measurements were performed on the {\LaFePO} with F concentrations of $x$ = 0, 0.03 and 0.05. The temperature ($T$) dependence of electrical resistivity $\rho(T)$ was measured by a standard four-probe method from room temperature down to 1.8 K. The electrical resistivity under magnetic fields ($H$) up to 14 T, $\rho_{xx}(T,H)$, was measured in the superconducting magnet. The specific heat $C(T)$ measurements were carried out by a relaxation technique from 20 K down to 1.8 K. The Hall coefficient $R_\mathrm{H}$ was measured in magnetic fields up to 5 T at various temperatures. 

%%%%%%%%%%%%%%%%%%%%%%%%%%%%%%%%%%%%%%%%%%%%%%%%%%%%%%%%%%
\section{Results and discussion}

\begin{figure}[tb]
	\begin{center}
		\includegraphics[width=120mm]{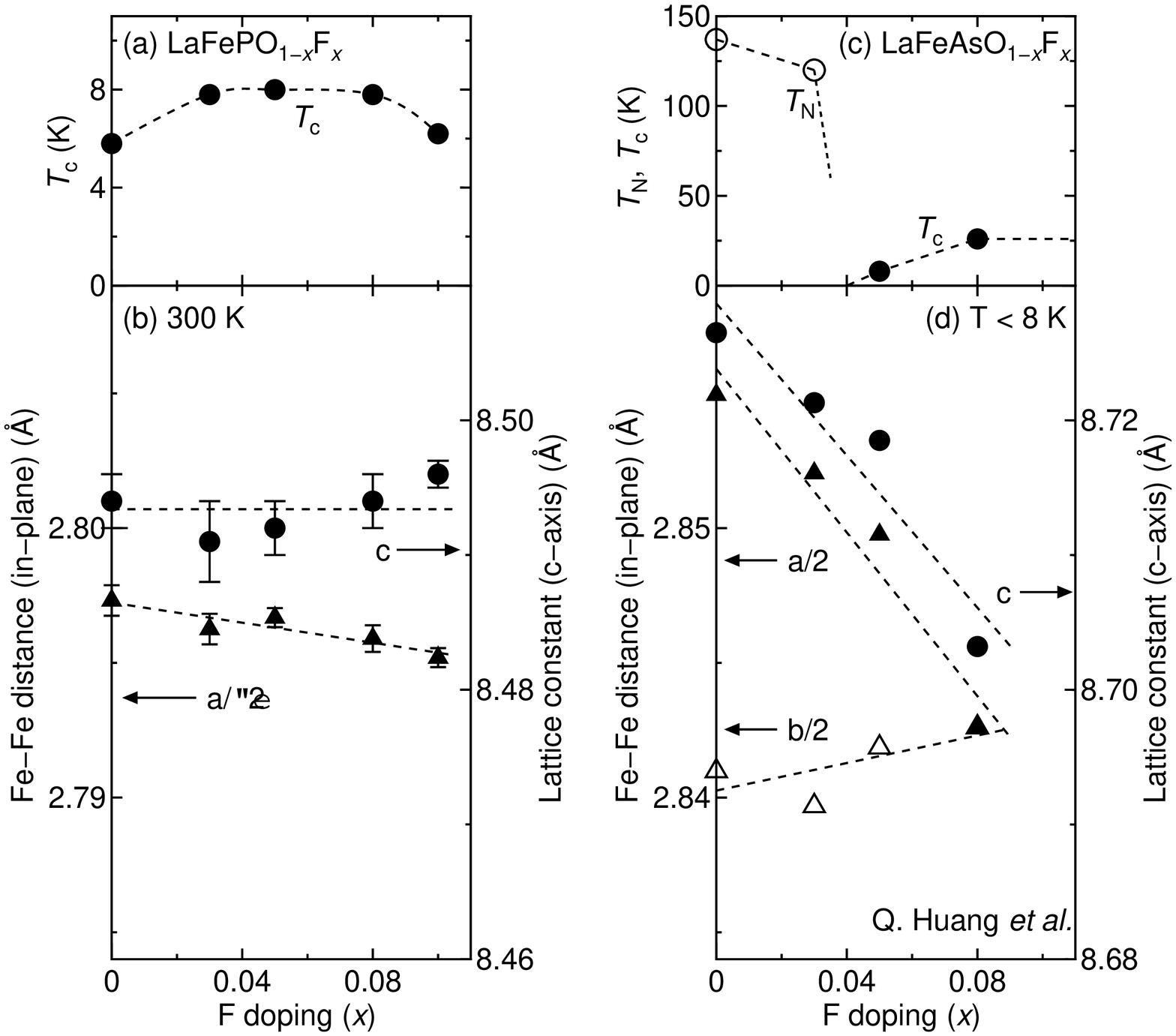}
	\end{center}
	\caption{F-doping ($x$) dependence of (a) transition temperatures for superconductivity ($T_\mathrm{c}$), and (b) lattice constants at 300 K of {\LaFePO}. The $x$-dependence of (c) $T_\mathrm{c}$ and N\'{e}el temperature $T_\mathrm{N}$, and (d) low temperature lattice constants for {\LaFeAsO} are also shown, which have been reported in ref. [11]. 
In upper panels ((a) and (c)), closed and open circles indicate $T_\mathrm{c}$ and $T_\mathrm{N}$, respectively. In lower panels ((b) and (d)), circles, closed and open triangles indicate the lattice constant of $c$-axis, Fe-Fe distance along $a$-axis and that of $b$-axis, respectively. Broken lines are the guide for eyes.}
	\label{fig:latconst}
\end{figure}

Figures~\ref{fig:latconst}(a) and (b) present the F-doping ($x$) dependence of {\Tc}, in-plane ($a$) and out-of-plane lattice constants ($c$) of {\LaFePO} with the tetragonal lattice. For comparison, the $x$-dependent {\Tc}, N\'{e}el temperature $T_\mathrm{N}$, and lattice constants of {\LaFeAsO} are plotted in Figs.~\ref{fig:latconst}(c) and (d), which were reported in ref. [11]. 
As shown in Fig. 1(b), the F-doping dependence of $c$ is small and not beyond an experimental error in this FeP system. In contrast, the $a$ monotonously decreases with increasing F concentration, and the decreasing ratio is about 0.6 \% per an atomic mole of F. The systematic change of $a$ is the evidence that F is successfully introduced into this FeP system.~\cite{errors} \par

In the FeAs system, both of in-plane and out-of plane lattice constants show a clear $x$-dependence. As shown Fig. 1(d), the $a$ and $c$ decrease with rates of approximately 5 \%  and 3.5 \% per F atomic mole, respectively, while the $b$-axis is slightly expanded with $x$. The changing rates of the lattice constants of the FeAs system are much larger than those in the FeP one. These results imply that the F doping controls only the band filling of the Fe 3d orbitals in {\LaFePO}, while in the FeAs system it changes not only the valence of Fe ions but also the crystallographic lattice coupled with electronic structure. \par

The different influence of F doping on the lattice structure seems to induce the different change on the electronic and magnetic features in these systems. As shown in Fig. 1(a), the {\Tc} shows a little but systematic dependence on the F-doping level in {\LaFePO}. The {\Tc} are $\sim$ 5.8, 7.8, 8.0, 7.8, and 6.2 K for $x$ = 0, 0.03, 0.05, 0.08, and 0.10, respectively. Only 3 \% F doping enhances {\Tc} by 2 K. Further doping hardly changes the value of {\Tc}. We synthesized several samples with various fluorine concentrations and oxygen vacancy. However, we have been not able to enhance {\Tc} further. 
In contrast, {\LaFeAsO} shows a drastic change of electronic and magnetic states induced by F doping, i.e. the phase transition from the antiferromagnetic metal to superconductor takes place at low temperatures. (See Fig. 1(c).) The {\Tc} and $T_\mathrm{N}$ show a systematic change against the F concentration $x$. \par
\begin{figure}[tb]
	\begin{center}
		\includegraphics[width=120mm]{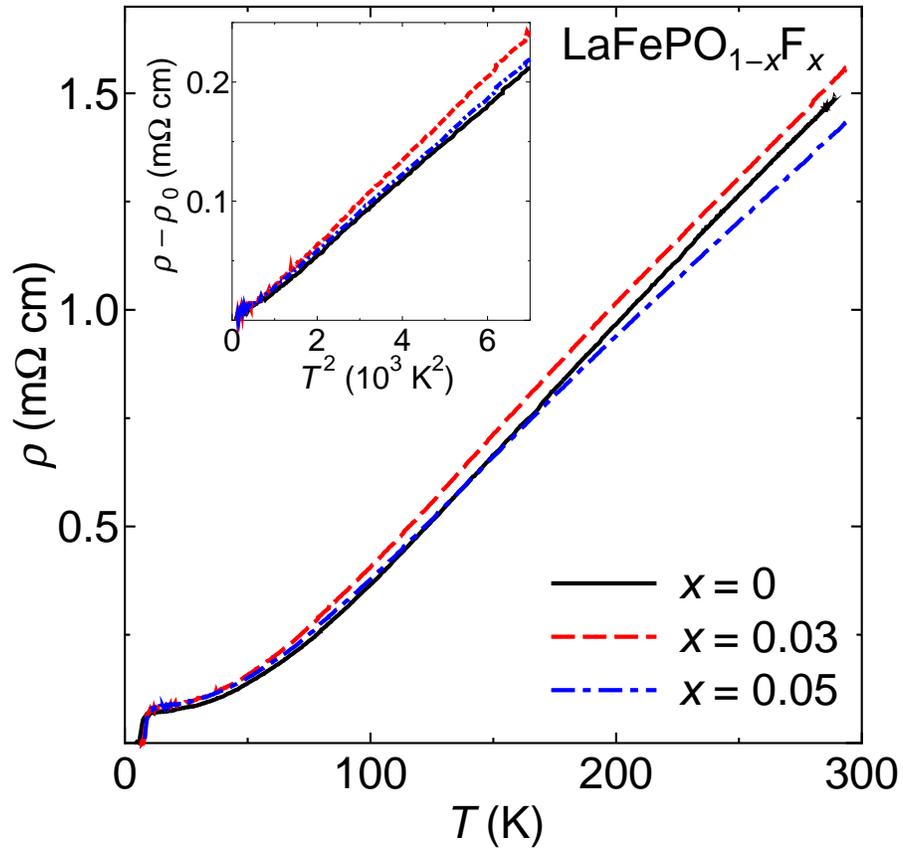}
	\end{center}
	\caption{(Color online) Temperature ($T$) dependence of electrical resistivity $\rho$ for {\LaFePO} with $x$ = 0 (black solid line), $x$ = 0.03 (red broken one) and $x$ = 0.05 (blue dash-dotted one). Inset shows plots of $\rho - \rho_0$ versus $T^2$, where $\rho_0$ is residual resistivity. The magnitude of $\rho$ just above {\Tc} is used as the $\rho_0$.}
	\label{fig:res}
\end{figure}

Next, we examine the effect of electron doping on transports and specific heat. The electrical resistivity $\rho(T)$ for $x$=0, 0.03 and 0.05 presented in Fig. \ref{fig:res} shows a metallic behavior and a phase transition to the superconducting state at low temperatures. The residual resistivity ratios ($RRR$) are about 20 in all the samples. 
Shown in the inset of Fig. \ref{fig:res}, $\rho(T)$ is proportional to $T^2$ at low temperatures. This $T^2$ dependence of $\rho(T)$ indicates the existence of the electron-electron correlation in the present system.~\cite{pines} The $T^2$ dependence of $\rho(T)$ has been also observed in {\LaFeAsO}.~\cite{sefat} 
In contrast to the case of FeAs system, the F-doping dependence of $\rho(T)$ is quite small in this FeP system. \par

\begin{figure}[tb]
	\begin{center}
		\includegraphics[width=120mm]{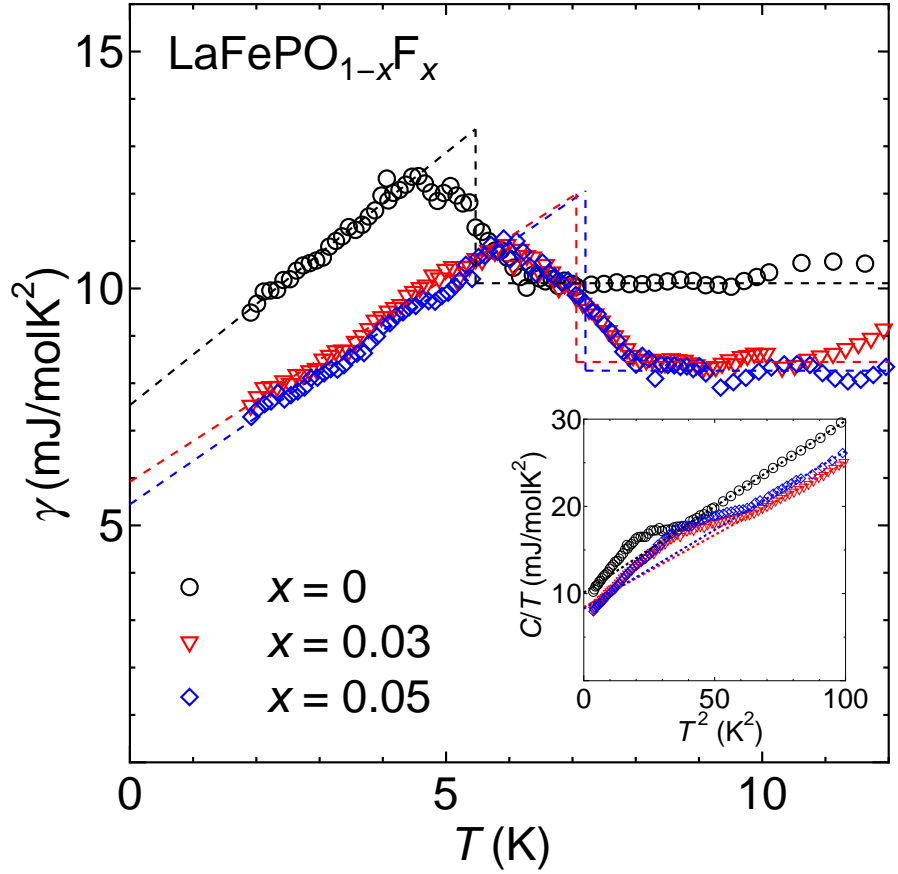}
	\end{center}
	\caption{(Color online) Temperature ($T$) dependence of electronic specific heat coefficient $\gamma$ for {\LaFePO} with $x$ = 0 (black circles), $x$ = 0.03 (red inverted triangles) and $x$ = 0.05 (blue diamonds). The broken lines show an ideal specific heat jump assuming a discontinuous entropy-conserved superconducting transition. Inset shows the plot of specific heat divided by temperature $C/T$ versus $T^2$. The broken line indicates the fitting results above {\Tc} by $C/T = \gamma_n + {\beta}T^2$, where $\gamma_n$ and $\beta$ are electronic and lattice specific heat coefficients.}
	\label{fig:sh}
\end{figure}

Temperature dependence of specific heat $C(T)$ for {\LaFePO} with the F concentrations of $x$=0, 0.03, and 0.05, plotted as $C/T$ vs. $T^2$, are shown in the inset of Fig. \ref{fig:sh}. Above {\Tc}, the data can be well fitted by $C/T = \gamma_n + \beta T^2$, where $\gamma_n$ and $\beta$ are electronic and lattice specific heat coefficients. $\gamma_n$ are 10.1, 8.5 and 8.3 mJ/molK$^2$ for $x$ = 0, 0.03 and 0.05, respectively.
Using calculated density of states at the Fermi level $g(E_\mathrm{F})$ of LaFePO, $\sim$ 2.5 states/eV per formula unit~\cite{lebegue}, $\gamma_n = (\pi^{2}/3)k_\mathrm{B}^{2}g(E_\mathrm{F})$ is estimated to be $\sim$ 5.9 mJ/molK$^2$. The obtained values are $\sim$ 1.5 times larger than the calculated one, which indicates the carrier mass enhancement due to the electron-electron correlation in {\LaFePO}. \par

The electronic part of $C/T$ in the normal and superconducting states, $\gamma (T)$, is estimated by subtracting the lattice contribution $\beta T^2$ from $C/T$, i.e. $\gamma (T) = C/T - \beta T^2$. 
$\gamma (T)$ in Fig. \ref{fig:sh} shows the specific heat jumps $\Delta C/T_\mathrm{c}$ near {\Tc}, suggesting that the samples are bulk superconductors. Temperature of the midpoint of the jump is consistent with the temperature where $\rho$ becomes zero. The normalized specific heat jumps $\Delta C/\gamma_n T_\mathrm{c}$ are 0.32, 0.42 and 0.46 for $x$ = 0, 0.03 and 0.05, respectively. 
These values are much smaller than that expected from the BCS weak-coupling limit ($\Delta C/\gamma_n T_\mathrm{c}$ = 1.43). 
(The jumps of specific heat around $T_\mathrm{c}$ are broad in the present result.~\cite{Tc} Assuming that the entropy change due to the superconducting transition occurs at $T_\mathrm{c}$, the values of $\Delta C/T_\mathrm{c}$ are still smaller than the theoretical one.) 
In addition, residual $\gamma$ is observed by extrapolating $\gamma(T)$ down to $T$ = 0. These behaviors of $\gamma(T)$ may be caused by Cooper pair breaking by the some impurities. Recently, theoretical study indicates that this FeP superconductor has a nodal superconducting gap.~\cite{kuroki1} The present results may support this proposal of the nodal superconducting gap for the FeP system. \par

\begin{figure}[tb]
	\begin{center}
		\includegraphics[width=120mm]{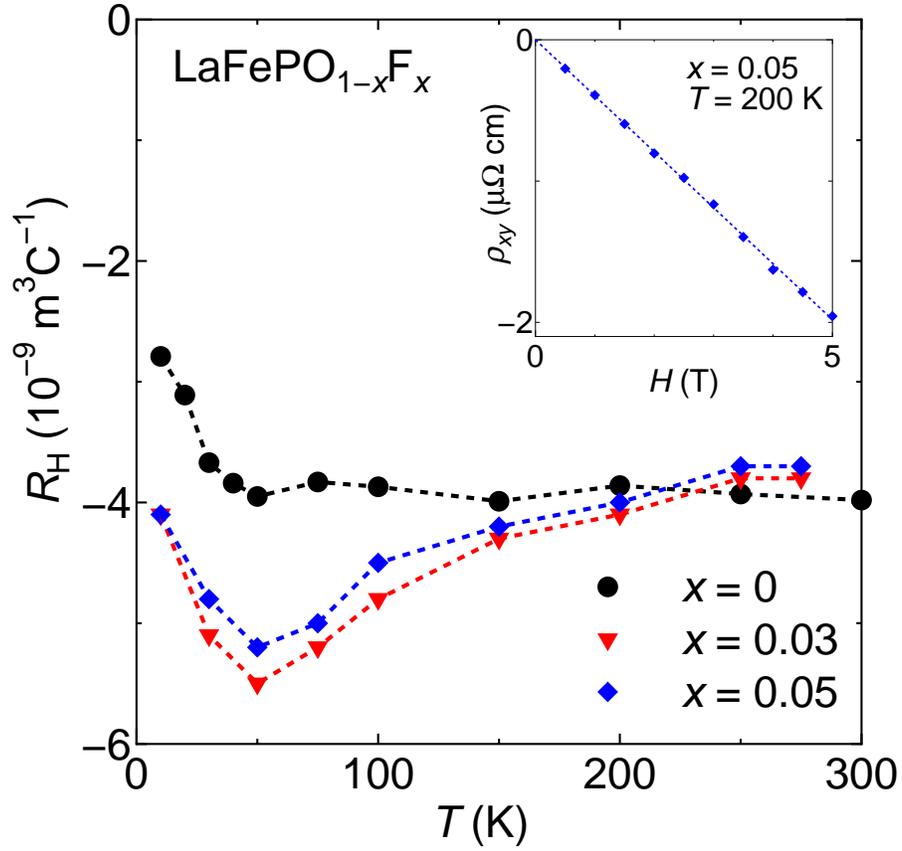}
	\end{center}
	\caption{(Color online) Temperature ($T$) dependence of Hall coefficient $R_\mathrm{H}$ for {\LaFePO} with $x$ = 0 (black circles), $x$ = 0.03 (red inverted triangles) and $x$ = 0.05 (blue diamonds). The broken lines are the guide for eyes. Inset shows a magnetic field $H$ dependence of Hall resistivity $\rho_{xy}$ at $T$ = 200 K for the sample with $x$ = 0.05. The broken line fits to the data by $\rho_{xy} = R_\mathrm{H}H$, indicating that the $\rho_{xy}$ is proportional to $H$ up to 5 T.}
	\label{fig:hall}
\end{figure}

Figure \ref{fig:hall} and the inset present the temperature-dependent Hall coefficients $R_\mathrm{H}$ for {\LaFePO} with $x$=0, 0.03, and 0.05, and the magnetic-field-dependent Hall resistivity $\rho_{xy}(T,H)$ for $x$=0.05 at 200 K. The Hall resistivity $\rho_{xy}(T,H)$ shown in the inset of Fig. \ref{fig:hall} is proportional to magnetic field $H$. The $\rho_{xy}(T,H)$ for all {\LaFePO} samples has the same $H$-dependence in the whole temperature region, indicating no anomalous Hall effect in this system. As shown in Fig. \ref{fig:hall}, Hall coefficients for all the samples are negative and have a large magunitude. Near room temperature, all three compounds have similar values, $R_\mathrm{H} \sim -4 \times 10^{-9}$ m$^3$/C. If a single band is assumed, $R_\mathrm{H} = 1/nq$, where $n$ is carrier concentration and $q$ is a carrier of a charge. Then $n$ at room temperature is estimated to be $\sim$ 0.1 electrons per unit formula, which indicates the low carrier density, similar to the case for the FeAs system.~\cite{sefat,jaroszynski} With decreasing temperature down to 50 K, $|R_\mathrm{H}|$ for {\LaFePO} with $x$=0.03 and 0.05 are enhanced, while that for undoped sample is independent of temperature. The magnitude of $R_\mathrm{H}$ for all samples decreases rapidly below 50 K. These temperature dependence of $R_\mathrm{H}$ is probably due to the existence of multiple carriers. In spite of electron doping by F substitution, $R_\mathrm{H}$ shows no clear change with increasing F content. It probably because the opposite changes of two carriers compensate with each other. \par

%It is probably because the parent compound LaFePO is already a good metal and thus a small amount of doping does not affect a carrier density so much, or because the opposite changes of two carriers compensate with each other. \par

\begin{figure}[tb]
	\begin{center}
		\includegraphics[width=160mm]{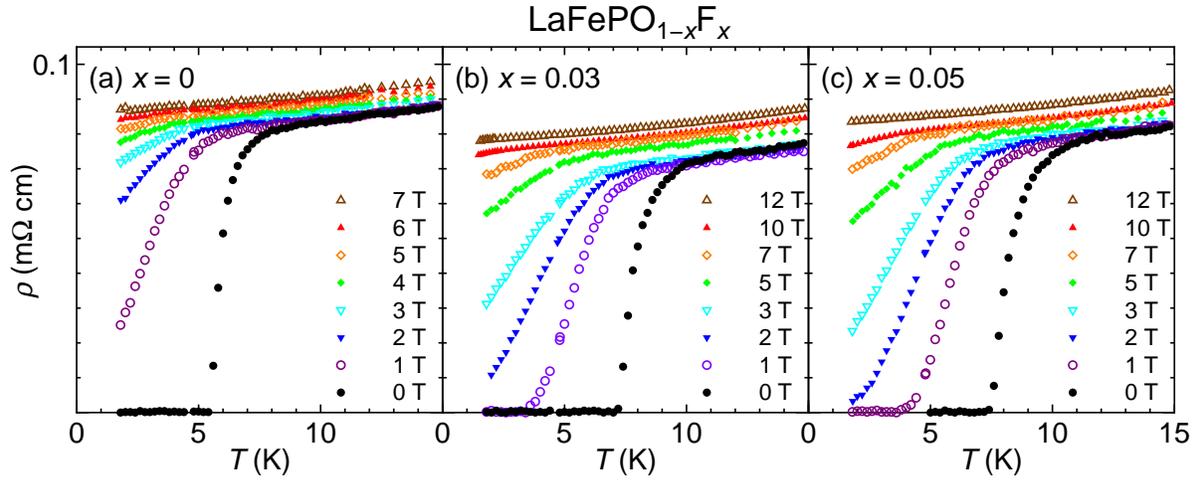}
	\end{center}
	\caption{(Color online) Temperature ($T$) dependence of resistivity $\rho$ in various magnetic fields for {\LaFePO} with (a) $x$ = 0, (b) $x$ = 0.03, and (c) $x$ = 0.05.}
	\label{fig:restrans}
\end{figure}

The temperature-dependent resistivity for {\LaFePO} with $x$ = 0, 0.03, and 0.05 in various magnetic fields are shown in Fig. \ref{fig:restrans}. The resistive transition between normal and superconducting states shifts to lower temperature, and the zero resistive state at the lowest temperature is not observed above 1, 2 and 3 T in $x$=0, 0.03 and 0.05 samples, respectively. The suppression of resistivity at low temperature completely disappears above 5 and 10 T in the pure and F-doped samples, respectively. 
The width of resistive transition broadens with increasing magnetic field $H$. 
This behavior is often observed in high-{\Tc} cuprates. The present results have indicated the large anisotropy of the upper critical field $H_\mathrm{c2}$, as expected from the 2-dimensional electronic structure. For the end member LaFePO, the anisotropic $H_\mathrm{c2}$ has been confirmed by the investigation using a single crystal.~\cite{hamlin} \par

\begin{figure}[tb]
	\begin{center}
		\includegraphics[width=155mm]{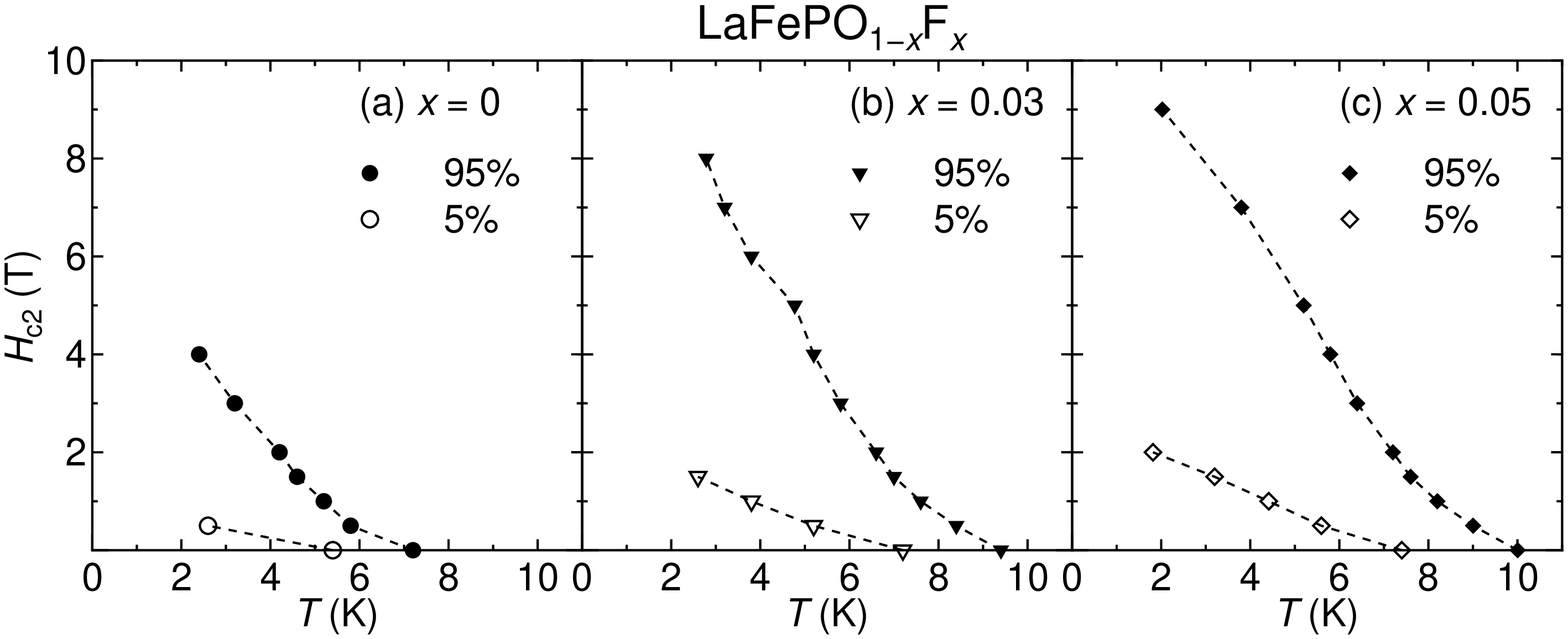}
	\end{center}
	\caption{Temperature ($T$) dependence of the upper critical fields $H_{c2}(T)$ for {\LaFePO} with (a) $x$ = 0, (b) $x$ = 0.03, and (c) $x$ = 0.05. Filled marks denote 95{\%} of the $\rho$ in the normal state and open ones 5{\%}. (See the main text for the definition.)}
	\label{fig:hc2}
\end{figure}

Figure \ref{fig:hc2} shows temperature dependence of the upper critical fields $H_\mathrm{c2}(T)$ for $x$ = 0, 0.03, and 0.05. The $H_\mathrm{c2}(T)$ indicated by the open and closed marks in Fig. \ref{fig:hc2} are defined as the magnetic fields, where $\rho_{xx}(T,H)$ drops to 95{\%} and 5{\%} of its extrapolated normal state value. Considering the anisotropy of $H_\mathrm{c2}$ described above and referring to the previous works for the FeAs polycrystalline samples~\cite{jaroszynski,hunte}, they are likely to correspond to the upper critical fields perpendicular to the $c$-axis $H_\mathrm{c2}^{ab}(T)$, and parallel to the $c$-axis $H_\mathrm{c2}^{c}(T)$, respectively. The upper critical fields at zero temperature are roughly estimated as $H_\mathrm{c2}^{ab}$ $\sim$ 7 T and $H_\mathrm{c2}^{c}$ $\sim$ 1 T for the undoped sample, which are consistent with the results by using the single crystal of LaFePO,~\cite{hamlin} and 12 and 2.5 T for the F-doped samples. The anisotropy of $H_\mathrm{c2}$ seems to be weakly reduced by F doping. The concave curvature around $T_\mathrm{c}$ is shown in $H_\mathrm{c2}$ for 95{\%} in Fig. \ref{fig:hc2}. It is probably the evidence for the two gap superconductivity in {\LaFePO}, as was observed in MgB$_2$.~\cite{eltsev} Recent study of thermal conductivity also indicates the multigap of superconductivity in the undoped system.~\cite{yamashita} \par

The present results of transports and specific heat measurements for this FeP system are suggestive of the existence of electron-electron correlations, the multi-bands, 2-dimensional electronic structure, and two superconducting gaps. These characters in the normal and superconducting states for the FeP system are similar to those for FeAs ones. 
In contrast to these similar electronic features, there is a distinct difference in F-doping dependence of {\Tc}, transport and magnetic properties. 
We think that this is related to the different lattice structural change by F doping between {\LaFePO} and {\LaFeAsO}, as shown in Fig. 1. 
In the iron-based pnictide superconductors, {\Tc} and other physical properties are very sensitive to the lattice distortion presumably because it modifies the band structure.
The present results may indicate that F-doping effect in the FeAs system is mainly induced by the change of lattice rather than by the change of the band filling. 
In other words, a pure electron-doping effect in the Fe-based pnictides is weak, as is observed in the FeP system. 
The crystal structure of {\LaFePO} is far from the optimum structure, for example the best Fe-P bond angle, for high-{\Tc} superconductivity.~\cite{lee} \par

%%%%%%%%%%%%%%%%%%%%%%%%%%%%%%%%%%%%%%%%%%%%%%%
\section{Summary}
We have synthesized polycrystalline {\LaFePO} with $x$ = 0-0.10, and systematically investigated the F-doping ($x$)-dependence of lattice parameters, electrical resistivity, specific heat, Hall coefficient and resistive transition in various magnetic fields. \par
At low temperatures, the existence of moderate electron-electron correlation is indicated from $T^2$ dependence of electrical resistivity and enhancement of electronic specific heat coefficient. A clear jump of electronic specific heat around {\Tc} indicates that the samples exhibit bulk superconductivity. Large magnitude and temperature dependence of Hall coefficient suggests the low carrier density and the existence of multiple carriers in this system. Temperature dependence of the upper critical field suggests that the system is two gap superconductor. The upper critical field at the lowest temperature perpendicular and parallel to the $c$-axis, $H_\mathrm{c2}^{ab}$ and $H_\mathrm{c2}^{c}$, are roughly estimated as 7 and 1 T for the undoped sample and 12 and 2.5 for the F-doped samples, respectively. \par 
The present results indicate that electron doping without large crystallographic change by F-substitution gives little effect on the electronic properties. 
This suggests that the observed change in {\Tc} and physical properties in the FeAs system is predominantly caused by the structural change rather by the electron doping due to F substitution. 
In the FeP system, F doping monotonically and slightly decreases the in-plane lattice constant. Although only 3 \% of fluorine doping enhances {\Tc} by 2 K, further doping barely change not only the {\Tc} but also other physical properties. 
%%%%%%%%%%%%%%%%%%%%%%%%%%%%%%%%%%%%%%%%%%%%%%%
\section*{Acknowledgment}
We would like to thank S. Saijo, T. Masui, K. Tanaka, S. Araki, T. Nakano, Y. Nozue, S. Iguchi, and Y. Tokura for technical support and helpful discussion.
%%%%%%%%%%%%%%%%%%%%%%%%%%%%%%%%%%%%%%%%%%%%%%%%

%%%%%%%%%%%%%%%%%%%%%%%%%%%%%%%%%%%%%%%%%%%%%%%%%%%%%%%%%%%%%%%%%%%%%%%%%%%%%%%
\clearpage

\clearpage

\begin{thebibliography}{99} %% The number "99" means that this list has more than nine items.

\bibitem{kamihara1} Y. Kamihara, H. Hiramatsu, M. Hirano, R. Kawamura, H. Yanagi, T. Kamiya, and H. Hosono: J. Am. Chem. Soc. \textbf{128} (2006) 10012.
\bibitem{watanabe} T. Watanabe, H. Yanagi, T. Kamiya, Y. Kamihara, H. Hiramatsu, M. Hirano, and H. Hosono: Inorg. Chem. \textbf{46} (2007) 7719.
\bibitem{kamihara0} Y. Kamihara, T. Watanabe, M. Hirano, and H. Hosono: J. Am. Chem. Soc. \textbf{130} (2008) 3296.
\bibitem{chen1} X. H. Chen, T. Wu, G. Wu, R. H. Liu, H. Chen, and D. F. Fang: Nature \textbf{453} (2008) 761.
\bibitem{ren1} Z. A. Ren, J. Yang, W. Lu, W. Yi, X. L. Shen, Z. C. Li, G. C. Che, X. L. Dong, L. L. Sun, F. Zhou, and Z. X. Zhao: Europhys. Lett. \textbf{82} (2008) 57002.
\bibitem{chen2} G. F. Chen, Z. Li, D. Wu, G. Li, W. Z. Hu, J. Dong, P. Zheng, J. L. Luo, and N.L. Wang: Phys. Rev. Lett. \textbf{100} (2008) 247002.
\bibitem{kito} H. Kito, H. Eisaki, and A. Iyo: J. Phys. Soc. Jpn. \textbf{77} (2008) 063707.
\bibitem{vildosola} V. Vildosola, L. Pourovskii, R. Arita, S. Biermann, and A. Georges: Phys. Rev. B \textbf{78} (2008) 064518.
\bibitem{cruz} C. de la Cruz, Q. Huang, J. W. Lynn, J. Y. Li, W. Ratcliff II, J. L. Zarestky, H. A. Mook, G. F. Chen, J. L. Luo, N. L. Wang, and P. C. Dai: Nature \textbf{453} (2008) 899.
\bibitem{klauss} H. H. Klauss, H. Luetkens, R. Klingeler, C. Hess, F. J. Litterst, M. Kraken, M. M. Korshunov, I. Eremin, S. L. Drechsler, R. Khasanov, A. Amato, J. Hamann-Borrero, N. Leps, A. Kondrat, G. Behr, J. Werner, and B. B\"{u}chner: Phys. Lev. Lett. \textbf{101} (2008) 077005.
\bibitem{huang} Q. Huang, J. Zhao, J. W. Lynn, G. F. Chen, J. L. Luo, N. L. Wang, and P. C. Dai: Phys. Rev. B \textbf{78} (2008) 054529.
\bibitem{luetkens} H. Luetkens, H. H. Klauss, M. Kraken, F. J. Litterst, T. Dellmann, R. Klingeler, C. Hess, R. Khasanov, A. Amato, C. Baines, M. Kosmala, O. J. Schumann, M. Braden, J. Hamann-Borrero, N. Leps, A. Kondrat, G. Behr, J. Werner, and B. B\"{u}chner: Nat. Mater. \textbf{8} (2009) 305.
\bibitem{kamihara2} Y. Kamihara, M. Hirano, H. Yanagi, T. Kamiya, Y. Saitoh, E. Ikenaga, K. Kobayashi, and H. Hosono: Phys. Rev. B \textbf{77} (2008) 214515.
\bibitem{liang} C. Y. Liang, R. C. Che, H. X. Yang, H. F. Tian, R. J. Xiao, J. B. Lu, R. Li, and J. Q. Li: Supercond. Sci. Technol. \textbf{20} (2007) 687.
\bibitem{kondrat} A. Kondrat, J. E. Hamann-Borrero, N. Leps, M. Kosmala, O. Schumann, J. Werner, G. Behr, M. Braden, R. Klingeler, B. B\"{u}chner, and C. Hess: cond-mat/0811.4436.
\bibitem{zimmer} B. I. Zimmer, W. Jeitschko, J. H. Albering, R. Glaum, and M. Reehuis: J. Alloys Compd. \textbf{229} (1995) 238.
\bibitem{errors} As shown in Fig. 1(b), the lattice constant of $a$ has an error due to inhomogeneity of F concentration $x$ in the samples. The errors are approximately 0.01$\sim$0.02 \% of lattice constant $a$. From these errors, it is estimated that the F concentration $x$ in LaFePO$_{1-x}$F$_x$ is distributed in the range of $\pm$0.02$\sim$0.03. Perhaps, the errors of lattice constants also include the instrumental ones, and the degrees of the actual inhomogeneity of F concentration are below the above-estimated value.
\bibitem{pines} D. Pines and Ph. Nozi\`{e}res: $The$ $Theory$ $of$ $Quantum$ $Liquids$ (W. A. Benjamin Inc., 1966).
\bibitem{sefat} A. S. Sefat, M. A. McGuire, B. C. Sales, R. Jin, J. Y. Howe, and D. Mandrus: Phys. Rev. B \textbf{77} (2008) 174503.
\bibitem{lebegue} S. Leb\`{e}gue: Phys. Rev. B \textbf{75} (2007) 035110.
\bibitem{Tc} From the broad jump of specific heat near $T_\mathrm{c}$, the superconducting transition of the present samples takes place at $T_\mathrm{c} \pm$ 0.8 $\sim$ 1.0 K. 
\bibitem{kuroki1} K. Kuroki, H. Usui, S. Onari, R. Arita, and H. Aoki: cond-mat/0904.2612. 
\bibitem{jaroszynski} J. Jaroszynski, S. C. Riggs, F. Hunte, A. Gurevich, D. C. Larbalestier, G. S. Boebinger, F. F. Balakirev, A. Migliori, Z. A. Ren, W. Lu, J. Yang, X. L. Shen, X. L. Dong, Z. X. Zhao, R. Jin, A. S. Sefat, M. A. McGuire, B. C. Sales, D. K. Christen, and D. Mandrus: Phys. Rev. B \textbf{78} (2008) 064511.
\bibitem{hamlin} J. J. Hamlin, R. E. Baumbach, D. A. Zocco, T. A. Sayles, and M. B. Maple: J. Phys.: Condens. Matter \textbf{20} (2008) 365220.
\bibitem{hunte} F. Hunte, J. Jaroszynski, A. Gurevich, D. C. Larbalestier, R. Jin, A. S. Sefat, M. A. McGuire, B. C. Sales, D. K. Christen, and D. Mandrus: Nature \textbf{453} (2008) 903.
\bibitem{eltsev} Y. Eltsev, S. Lee, K. Nakano, N. Chikumoto, S. Tajima, N. Koshizuka, and M. Murakami: Phys. Rev. B \textbf{65} (2002) 140501.
\bibitem{yamashita} M. Yamashita, N. Nakata, Y. Senshu, S. Tonegawa, K. Ikada, K. Hashimoto, H. Sugawara, T. Shibauchi, and Y. Matsuda: cond-mat/0906.0622. 
\bibitem{lee} C.-H. Lee, A. Iyo, H. Eisaki, H. Kito, M. T. Fernandez-Diaz, T. Ito, K. Kihou, H. Matsuhata, M. Braden, and K. Yamada: J. Phys. Soc. Jpn. \textbf{77} (2008) 0837047.

\end{thebibliography}
\end{document}